# Deep UV plasmonic enhancement of single protein autofluorescence in zero-mode waveguides

Aleksandr Barulin, Jean-Benoît Claude, Satyajit Patra, Nicolas Bonod, Jérôme Wenger*

*Aix Marseille Univ, CNRS, Centrale Marseille, Institut Fresnel, 13013 Marseille, France*

* Corresponding author: jerome.wenger@fresnel.fr

**Abstract**

Single molecule detection provides detailed information about molecular structures and functions, but it generally requires the presence of a fluorescent marker which can interfere with the activity of the target molecule or complicate the sample production. Detecting a single protein with its natural UV autofluorescence is an attractive approach to avoid all the issues related to fluorescence labelling. However, the UV autofluorescence signal from a single protein is generally extremely weak. Here, we use aluminum plasmonics to enhance the tryptophan autofluorescence emission of single proteins in the UV range. Zero-mode waveguides nanoapertures enable observing the UV fluorescence of single label-free β-galactosidase proteins with increased brightness, microsecond transit times and operation at micromolar concentrations. We demonstrate quantitative measurements of the local concentration, diffusion coefficient and hydrodynamic radius of the label-free protein over a broad range of zero-mode waveguide diameters. While the plasmonic fluorescence enhancement has generated a tremendous interest in the visible and near-infrared parts of the spectrum, this work pushes further the limits of plasmonic-enhanced single molecule detection into the UV range and constitutes a major step forward in our ability to interrogate single proteins in their native state at physiological concentrations.

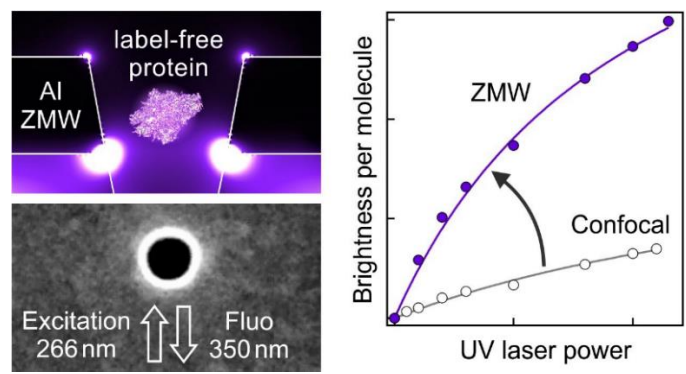

Figure for Table of Contents

Understanding proteins functions and structures requires investigations at the single molecule level to reveal dynamics and heterogeneities hidden in ensemble-averaged measurements.[1–3] Various approaches can provide a single molecule sensitivity,[4] and among them fluorescence detection is by far the most widely used method.[5,6] Superresolution imaging, Förster resonance energy transfer (FRET) and fluorescence correlation spectroscopy (FCS) are routinely implemented in laboratories. However, all these techniques are bound by the requirement of grafting an external fluorescent dye to the protein of interest. This fluorescence labelling can become a severe issue as fluorescent markers may perturb the protein reactivity or alter its conformation dynamics.[7–10] Moreover, fluorescence labelling is also a complex and time-consuming task, especially in the case of protein purification.

The three aromatic amino acid residues, tryptophan, tyrosine and phenylalanine naturally present in a vast majority of proteins, emit fluorescence light when excited in the 260-300 nm ultraviolet (UV) range.[6,11] Using this intrinsic protein autofluorescence rules out all the issues related with the external fluorescence labelling. While broadly used at the molecular ensemble level in fluorimeters and assays, the UV autofluorescence detection remains very challenging at the single molecule level. As compared to conventional fluorescent dyes, the aromatic amino acids feature lower absorption cross-sections, lower quantum yields and lower photostabilities.[6] Different excitation schemes using either one,[12–15] two,[16,17] or three[18] photon excitation have been explored, but single protein detection experiments based on UV autofluorescence remain scarce and limited by low signal to noise ratios. Let us also point out that the UV autofluorescence naturally occurs in all proteins containing tryptophan or tyrosine residues (more than 90% of all known proteins) and should not be confused with the visible-light intrinsic fluorescence of proteins related to the green fluorescent protein family (GFP, YFP, mCherry…).[19]

Plasmonic optical nanoantennas provide powerful means to overcome the limits of diffraction-limited microscopes, as they enable concentrating light at the nanoscale and enhancing the fluorescence emission of single molecules.[20–24] Additionally, plasmonic nanostructures offer the opportunity to detect single molecules at high micromolar concentrations compatible with biologically-relevant conditions.[25–27] All these features are highly appealing to improve the tryptophan autofluorescence detection of single proteins and extend plasmonics into the UV range.[28–30] While numerical simulations of plasmonic resonances in aluminum nanoparticles[31,32] and nanoapertures[33–35] predict single emitter fluorescence enhancement, the experiments on UV plasmonics have remained largely focused on dense molecular layers deposited on nanoparticle arrays to enhance Raman scattering,[36–39] or fluorescence.[40–43] Metal nanoapertures were shown to reduce the fluorescence lifetime,[44,45] but there has been no report so far quantifying the plasmonic-enhanced UV fluorescence at the single protein level. Several reasons contribute to make this a technical challenge: the need for reproducible well-controlled plasmonic structures, the need for a proper characterization approach to deal with the low brightness and photostability of proteins, and the limited stability of aluminum structures in water environment.[46–49]

Here, we report the label-free detection of single β-galactosidase proteins using their natural tryptophan UV fluorescence emission enhanced by plasmonic aluminum nanoapertures. Our experiments are rationally designed to overcome all the previous limitations. To provide reproducible and well-controlled aluminum plasmonic nanostructures, we use zero-mode waveguides (ZMW), which are nanoapertures of 35 to 150 nm diameters milled in opaque aluminum films.[25,50] These ZMWs enhance the fluorescence brightness and isolate single proteins in a concentrated micromolar solution. With their simple geometry and reproducible fabrication, ZMWs are well suited to serve as benchmark platforms to test the UV plasmonics fluorescence enhancement. Contrarily to gap antennas,[23,24] ZMWs feature a better defined detection volume and are less prone to dispersion due to fabrication defects. Contrarily to nanoparticles,[22] ZMWs allow to work on a quasi-dark background, ensuring low noise in the measurements.

The restricted UV photostability of proteins has been long recognized as a major limiting issue.[14,16,18] Here, we use oxygen scavengers[51] and reducing agents[52] to promote the tryptophan photostability. Additionally, the reduced diffusion time experienced by proteins in ZMWs helps to prevent observing the negative effects of photobleaching. While the corrosion of aluminum in water-based environments can be a severe issue for UV plasmonics,[46–48] we have recently developed a protocol to inhibit the aluminum photocorrosion under UV illumination.[49] Finally, we implement fluorescence correlation spectroscopy (FCS) as a robust analysis method to count the number of molecules and measure their diffusion time even in the presence of noisy fluorescence traces.[53,54] With all these elements taken together, we can demonstrate the label-free detection of single proteins using their UV autofluorescence enhanced by a plasmonic structure. Zero-mode waveguides, nanopores and nanoapertures have received a large interest for many biophysics applications including molecular sensing,[55–58] DNA sequencing,[59–61] enzymatic reaction monitoring,[62,63] and biomembrane investigations.[64,65] With their ability to probe the protein tryptophan autofluorescence demonstrated here, new possibilities are offered to interrogate single proteins in their native state at physiological concentrations.

Figure 1a presents the scheme of our experiment: a single ZMW milled in a 50 nm thick opaque aluminum film is positioned at the focus of a UV confocal microscope (the complete setup is detailed in the Supporting Information Fig. S1). The ZMW is covered with the solution containing the label-free proteins which freely diffuse across the nanoaperture volume. This geometry allows to record only the fluorescence stemming from the ZMW, whose volume can be three orders of magnitude lower than the diffraction-limited confocal volume.[25] Extending the ZMW optical response into the 200-400 nm UV range requires the use of aluminum instead of gold or silver, as these classical plasmonic metals feature too high losses below 400 nm.[28,30] The deposition of aluminum demands specific conditions of high deposition rates (>10nm/s) and low vacuum (<10$^{-6}$ mbar) to reduce the amount of residual oxide found within the bulk of the aluminum layer and ensure the best plasmonic performance.[28,66] Focused ion beam (FIB) milling then directly carves the nanoaperture into the aluminum layer, enabling an accurate control on the diameter and the 50 nm undercut in the quartz substrate to optimize the signal to noise ratio.[67–69] Figure 1b shows typical SEM images of our ZMW samples with different diameters.

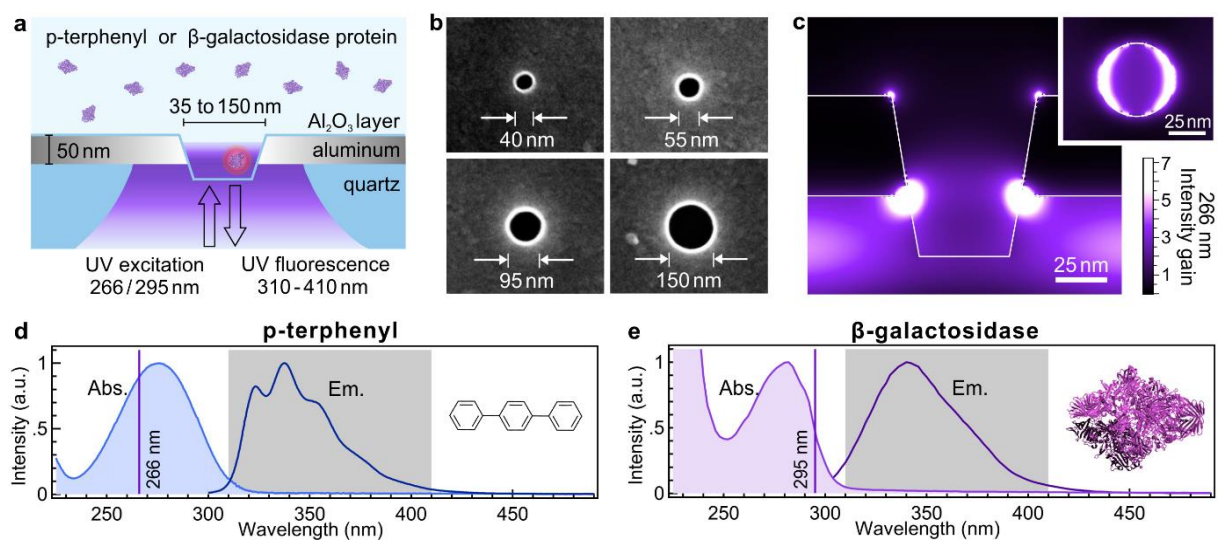

**Figure 1.** Aluminum zero-mode waveguides (ZMWs) to enhance single molecule fluorescence in the UV range. (a) Schematic view of the ZMW configuration. (b) Scanning electron microscope images of ZMWs with different diameters. (c) Finite difference time domain (FDTD) simulation of the intensity enhancement inside a 50 nm diameter aluminum ZMW at a 266 nm wavelength. The intensity is displayed along the laser linear polarization direction and the inset shows the intensity in a plane 5 nm inside the nanoaperture from the aluminum-quartz interface. (d) Absorption and emission spectra of p-terphenyl molecules in cyclohexane. (e) Absorption and emission spectra of β-galactosidase protein in aqueous buffer. In (d) and (e) the violet lines correspond to the laser wavelength used for excitation, and the grey backgrounds show the detection spectral range.

Circular apertures milled in a perfect electrical conductor have a theoretical cut-off diameter given by $0.59\,\lambda/n$ where $n$ is the refractive index of the medium filling the aperture.[70] Diameters below this cut-off lead to an evanescently decaying intensity profile as light is not transmitted anymore. For water-filled ZMWs at 266 nm, the theoretical cut-off diameter amounts to 115 nm. For a real metal of finite conductivity, the propagation constant inside the aperture differs slightly as part of the electromagnetic field penetrates into the metal (Supporting information Fig. S2). Taking this effect into account, our numerical simulations of the UV intensity profile inside a 50 nm ZMW display the characteristic features of evanescent decay and local intensity enhancement (Fig. 1c).

The β-galactosidase protein from *Escherichia coli* selected for this work has a tetrameric structure containing a total of 156 tryptophan residues.[14] Each of the four subunits has a length of 1024 aminoacids and a mass of 116.5 kDa. The quantum yield of tryptophan in water is 13%, but because of the fluorescence quenching occurring between nearby aminoacids, the net quantum yield of emission for tryptophan in proteins can be significantly lower.[6,71] Part of this quenching can be compensated by selecting a protein with a large number of tryptophan residues.[14,16] The main focus of this work is to assess the detectability of proteins inside plasmonic ZMWs. Because of the low signal to noise ratio inherent to protein UV autofluorescence, we first perform calibration measurements of the ZMW properties using p-terphenyl as a high quantum yield fluorescent dye in the UV range. This calibration is important to unambiguously confirm the validity of the results obtained on β-galactosidase proteins. p-terphenyl has a quantum yield of 93% in cyclohexane,[72] and features absorption and emission spectra very close to the ones of β-galactosidase, well into the UV range (Fig. 1d,e)

FCS and fluorescence lifetime experiments on p-terphenyl molecules assess the optical performance of ZMWs in the UV (Fig. 2). The choice of a high quantum yield UV dye enables an accurate calibration of the ZMW photonic properties to benchmark the protein autofluorescence that will be studied later in Fig. 3. Figure 2a displays the raw fluorescence intensity time traces recorded on ZMWs with different diameters (the background noise is negligible here). Throughout these experiments, the p-terphenyl concentration is kept constant at 10 μM. Each time trace is analyzed by FCS to compute the temporal correlation function and assess the average number of molecules $N_{mol}$ present in the ZMW detection volume (Fig. 2d) and their average diffusion time $\tau_d$ (Fig 2e). Figure 2b shows normalized FCS correlation functions to evidence the shorter diffusion times observed as the ZMW diameter is reduced. Details about the FCS fitting procedure are given in the Methods section, and supplementary FCS traces are shown in the Supporting Information Fig. S3 to better present the quality of the fits.

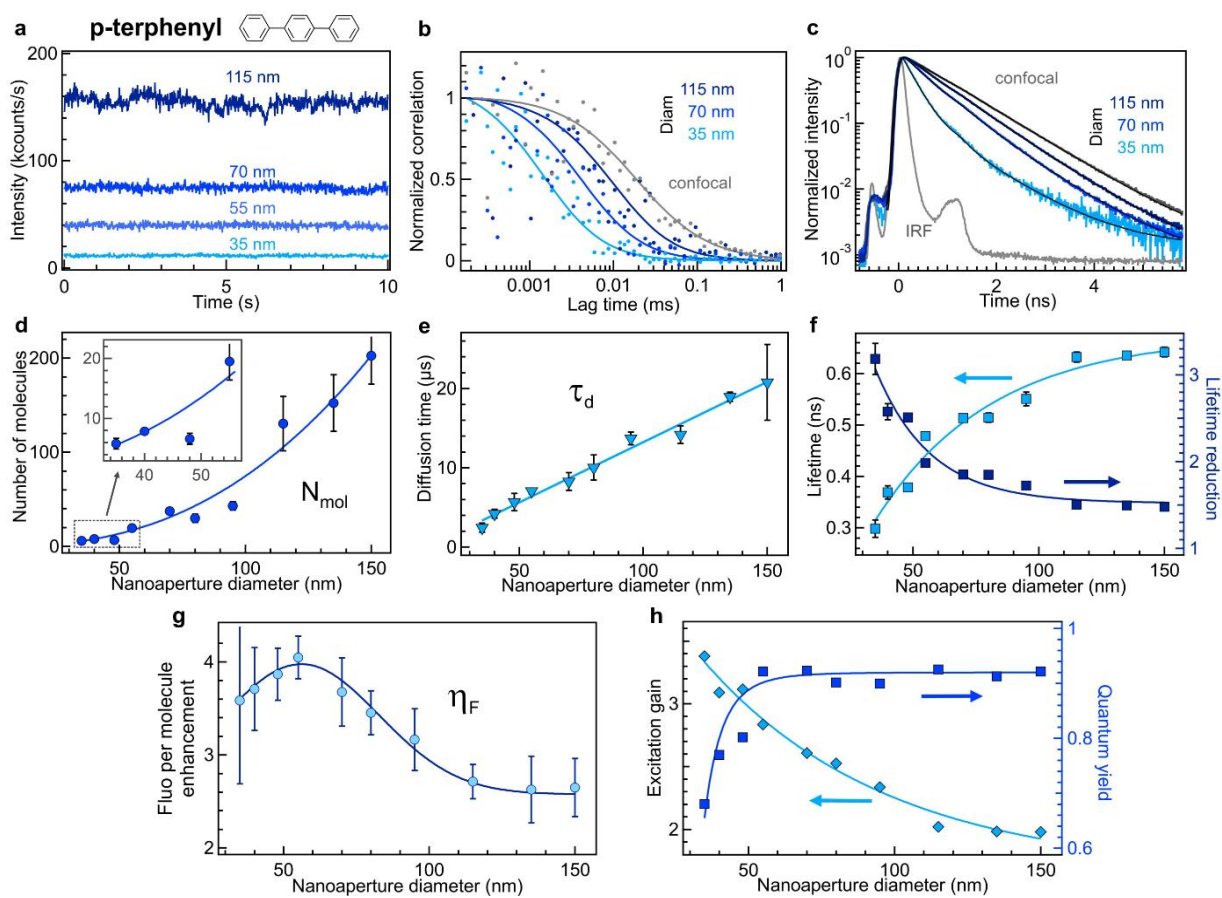

**Figure 2.** Characterization of aluminum ZMWs performance in the UV. (a) p-terphenyl fluorescence intensity time traces recorded on ZMWs with different diameters. For all the experiments shown here, the p-terphenyl concentration is constant at 10 µM. (b) Normalized FCS correlation functions showing a reduction of the diffusion time as the ZMW diameter is decreased. (c) Normalized fluorescence lifetime decay traces for the confocal reference and ZMWs of decreasing diameters. The black lines are numerical fits, and IRF indicates the instrument response function. (d) Number of p-terphenyl molecules in the ZMW detection volume and (e) average diffusion time deduced from the FCS fits as a function of the ZMW diameter. (f) p-terphenyl fluorescence lifetime measured from the decays in (c) and corresponding lifetime reduction as compared to the confocal reference. (g) Fluorescence brightness per molecule enhancement as a function of the ZMW diameter. (h) Quantum yield and excitation gain in ZMWs deduced from the lifetime reduction in (f) and the fluorescence enhancement in (g) with respect to the diameter. Throughout (d-h) the lines are guide to the eyes.

Less than 5 p-terphenyl molecules are detected in the 35 nm ZMW, while at the 10 µM concentration, the diffraction-limited confocal volume of 0.5 fL contains 3000 molecules. We find that $N_{mol}$ has a nearly cubic dependence with the ZMW diameter to the power 2.7 (Fig. 2d, note that this number of molecules also considers the undercut volume below the ZMW). The diffusion time $\tau_d$ scales linearly with the ZMW diameter (Fig. 2e), and a quadratic dependence with the intensity decay length inside the ZMW can be retrieved (Supporting Information Fig. S4), confirming the Brownian nature of the diffusion process. Additional control experiments assess the validity of the ZMW calibration data: we check that the number of detected molecules scales linearly with the concentration (Supporting Information Fig. S5) and that the fluorescence brightness per molecule depends linearly with the UV excitation power (Supporting Information Fig. S6). In addition to FCS, time-resolved fluorescence decay histograms record the fluorescence lifetime in the ZMWs and its reduction as compared to the 0.95 ns reference in the confocal setup (Fig. 2c,f). As the ZMW diameter is reduced, the dye gets in closer proximity to the aluminum surface leading to a 3× reduction of its fluorescence lifetime down to 0.3 ns for the smallest 35 nm ZMW (Fig. 2f and Supporting Information Tab. S1).

Putting together all the characterizations (Fig. 2a-f), we derive the ZMW influence on the UV fluorescence photodynamics (Fig. 2g,h). First, the fluorescence brightness per molecule is assessed by normalizing the total signal by the number of p-terphenyl molecules detected by FCS. Typically at 200 µW excitation power, the fluorescence brightness amounts to 1.9 kcounts/s per molecule in a 55 nm diameter ZMW while in the confocal setup it is only 0.45 kcounts/s (Supporting Information Fig. S6), indicating a 4× fluorescence enhancement for the 55 nm ZMW. The fluorescence enhancement follows a Gaussian distribution with the ZMW diameter (Fig. 2g), with a clear optimum around 60 nm. This size corresponds to the diameter where the real part of the propagation constant inside the ZMW vanishes and the light group velocity is minimum (Supporting Information Fig. S2).[73] To reveal the physics behind this phenomenon, we express the fluorescence enhancement $\eta_F$ as the product of the gains in the excitation intensity $\eta_{exc}$ times the gain in quantum yield $\eta_\phi$ and the gain in collection efficiency $\eta_{coll}$.[74] The quantum yield gain can be further written as the ratio between the gains in the radiative rate $\eta_{\Gamma rad}$ and the total decay rate $\eta_{\Gamma tot}$. This expresses the fluorescence enhancement as $\eta_F = \eta_{exc}\,\eta_{coll}\,\eta_{\Gamma rad}\,/\,\eta_{\Gamma tot}$ (with these notations $\eta_{\Gamma tot}$ corresponds also to the fluorescence lifetime reduction). In the case of a slightly focused laser beam (as with our 0.6 NA), it can be derived from the reciprocity theorem that the gain in excitation intensity amounts to the products of the gains in collection efficiency times the gain in radiative rate: $\eta_{exc} = \eta_{coll}\,\eta_{\Gamma rad}$.[75] Thanks to this simplification, we can now compute from the data in Fig. 2f,g the influence of the ZMW on the excitation intensity and the quantum yield (Fig. 2h). The excitation intensity increases up to 3.5× as the ZMW diameter is reduced, in good quantitative agreement with FDTD simulations. While the quantum yield does not depend much on the ZMW diameter for sizes above 60 nm (the radiative emission dominates the photokinetics pathways), a clear quenching is observed for the smallest apertures where nonradiative plasmonic losses become important. The trade-off between excitation gain maximization and quantum yield quenching minimization explains the optimum ZMW diameter around 60 nm.

Overall, Figure 2 completely characterizes the ZMW performance in the UV range, demonstrating excellent control, reproducibility and tunability of their photonic properties. This makes ZMW useful platforms for single molecule analysis in the UV, and is a significant step forward to extend plasmonics into the UV band. We can now apply this knowledge to demonstrate the label-free detection of tryptophan-containing proteins enhanced by UV plasmonics.

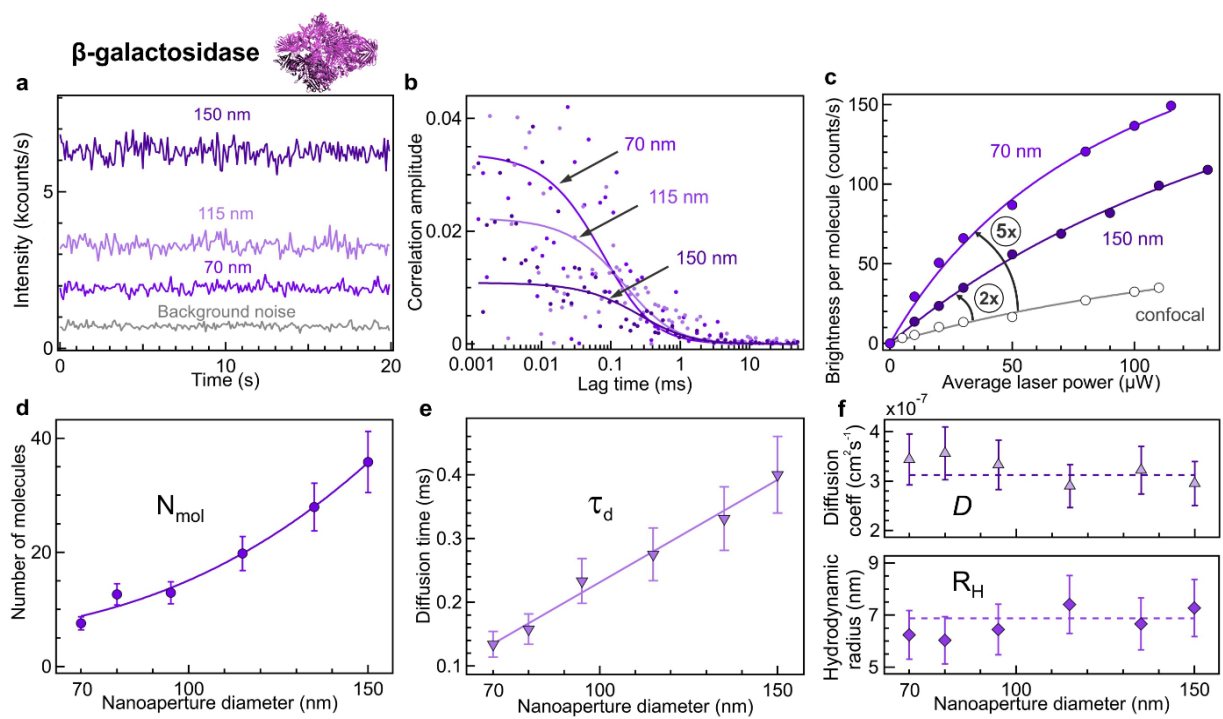

**Figure 3.** Label-free tryptophan autofluorescence detection of β-galactosidase proteins in aluminum ZMWs. (a) Fluorescence intensity time traces recorded with different ZMW diameters. For all the experiments shown here, the β-galactosidase protein concentration is 2 µM. (b) Raw FCS correlation functions corresponding to the traces shown in (a). Detailed FCS fitting results and residuals are shown in the Supporting Information Fig. S7. (c) Fluorescence brightness per molecule as a function of the average laser power, after the background noise has been subtracted. The lines are numerical fits based on a two-level fluorescence system. As compared to the confocal reference, the values for the 150 nm and 70 nm diameter ZMWs are enhanced by 2× and 5× respectively. (d) Number of β-galactosidase molecules in the ZMW detection volume and (e) average diffusion time as a function of the ZMW diameter deduced from the FCS fits. (f) From the diffusion time in (e) and using the experimental results obtained with p-terphenyl to calibrate the nanoaperture influence (Fig. 2e), we can retrieve experimentally the diffusion coefficient and hydrodynamic radius of β-galactosidase in ZMWs. The dashed horizontal lines indicate the values from the literature.[14]

Figure 3 summarizes our main results on the enhanced detection of β-galactosidase UV autofluorescence in ZMWs. Raw fluorescence intensity time traces are shown in Fig. 3a for ZMWs of different diameters. As compared to the p-terphenyl fluorescence and despite the 156 tryptophans contained in each protein, the detected intensity for β-galactosidase is about one order of magnitude lower, and the signal to background becomes a limiting issue. We could only record relevant fluorescence data for ZMWs diameters above 70 nm. For lower diameters, the data becomes too noisy as the signal to background goes below one. Importantly, our autofluorescence traces are stable over several minutes, and show no signs of photobleaching or photodegradation.[14] The fluorescence time traces are analyzed by FCS using the same procedure, with raw (non-normalized) FCS correlation traces displayed in Fig. 2b and detailed in the Supporting Information Fig. S7. Importantly, we check that the background noise does not yield any correlation (Supporting Information Fig. S8), confirming that the observed FCS correlations stem from the β-galactosidase proteins. The FCS analysis then quantifies the number of molecules $N_{mol}$ (Fig. 3d) and the diffusion time $\tau_d$ (Fig 3e) as a function of the ZMW diameter. The evolutions of $N_{mol}$ and $\tau_d$ follow exactly the same behavior with the ZMW diameter observed with p-terphenyl (Fig. 2d,e), confirming the validity of our results. From the known concentration of each molecular sample, we compute back the ZMW detection volume for each experiment. The values found for both β-galactosidase and p-terphenyl nicely coincide (Supporting Information Fig. S9), and establish our ability to count individual label-free proteins in ZMWs.

We use the number of molecules assessed by FCS to quantify the fluorescence brightness per β-galactosidase protein for the different ZMWs (Fig. 2c). All samples follow a fluorescence saturation curve with the excitation power, verifying that the detected signal stems from the protein tryptophan autofluorescence and not from the background or Raman scattering. At 100 µW excitation power, the fluorescence brightness is 140 counts/s per protein in a 70 nm diameter ZMW. This value is about 5× greater than the 30 counts/s per protein found for the confocal reference (Fig. 2c), demonstrating the occurrence of plasmonic fluorescence enhancement for the native tryptophan emission. However, this low intensity level prevents from directly observing UV fluorescence bursts clearly resolved in time, and requires the use of FCS as single molecule analysis tool.[19,53] The very weak autofluorescence signal in the confocal setup also highlights the interest for the plasmonic enhancement. Even moderate enhancement factors can make a significant contribution improving the detectability of single proteins above the background noise level. As for p-terphenyl, the fluorescence enhancement factor depends on the ZMW diameter, with an optimum size around 70 nm (Supporting Information Fig. S10).

The photostability of tryptophan autofluorescence in the UV has been recognized as a major limiting issue.[14,16,18] For the detection of single molecules in solution, the photobleaching probability depends on the time spent by the molecules in their excited state, and is proportional to the product of the excitation power times the diffusion time and the fluorescence lifetime.[76,77] Therefore, in the case of the ZMW, the effect of the higher excitation power (leading to a more pronounced photobleaching) is fortunately compensated by the shorter diffusion time across the ZMW. We evidence this experimentally as the excitation power dependence of the fluorescence intensity (Fig. 2c) indicates saturation, but not the drop characteristic of photobleaching.[76,77] Also the fluorescence time traces (Fig. 2a) are stable with time. Although we did not monitor any significant reduction of the β-galactosidase fluorescence lifetime in the ZMWs (Supporting Information Fig. S11, the photokinetics are too much dominated by the internal nonradiative conversion), the reduced diffusion time is by itself enough to stabilize the emission process and avoid observing photobleaching. This illustrates another benefit brought by the ZMWs. Furthermore, working with proteins featuring lower internal nonradiative rates would further promote the photostability in plasmonic nanostructures thanks to the additional fluorescence lifetime reduction.[78]

Finally, we show that quantitative measurements of the diffusion coefficients and hydrodynamic radii of label-free proteins can be performed in ZMWs. An accurate description of the diffusion behavior inside a ZMW is not even needed, as relative measurements using the p-terphenyl calibration efficiently circumvent this supplementary difficulty. As the FCS diffusion time is inversely proportional to the diffusion coefficient $D$, the ratio of the FCS diffusion times found for p-terphenyl and β-galactosidase determines the ratio of the diffusion coefficients: $\tau_{d,pter} / \tau_{d,\beta gal} = D_{\beta gal} / D_{pter}$. Using the value calibrated for p-terphenyl on the confocal UV microscope $D_{pter}$ = 5.6 $10^{-6}$ cm²/s, we can compute the diffusion coefficient for β-galactosidase in each ZMW (Fig. 3f). Additionally, the hydrodynamic radius $R_H$ can be deduced from the diffusion coefficient by using the Stokes-Einstein equation $D = kT/6\pi\eta R_H$ where $kT$ is the thermal energy and $\eta$ the water viscosity (Fig. 3f). For all the different ZMW diameters, we find that the experimental values for $D$ and $R_H$ correspond well to the 3.1 $10^{-7}$ cm²/s and 7 nm values reported in the literature.[14] This demonstrates that complete quantitative FCS measurements are possible on label-free proteins inside ZMWs. This agreement also further confirms that the protein structure is not damaged by the UV illumination inside the ZMW.

To conclude, this work underscores the high potential of aluminum plasmonics to enable UV autofluorescence studies of single proteins in their natural state. Combining deep UV plasmonics with the detection of single label-free proteins requires dedicated strategies to optimize the UV aluminum nanostructures fabrication, counteract the metal photocorrosion, deal with the limited photostability of proteins and develop robust analysis tools to extract useful information out of noisy traces. Here, we conclusively report the first demonstration of single protein tryptophan autofluorescence detection enhanced by UV aluminum plasmonics. Our specifically designed aluminum ZMWs enable observing the tryptophan fluorescence of label-free β-galactosidase proteins with single molecule sensitivity at micromolar concentrations, increased brightness per molecule and microsecond transit times. Quantitative FCS measurements are demonstrated over a wide range of ZMW diameters to measure the local concentration, diffusion coefficient and hydrodynamic radius of a label-free protein. This approach circumvents simultaneously the two main limitations of fluorescence labelling and dilutions to nanomolar concentrations that restrain confocal single molecule fluorescence detection. This novel facet of plasmonics constitutes an important step forward in our ability to interrogate single proteins in their native state at physiological concentrations. Currently, the detection sensitivity with the ZMW technique requires about 100 tryptophan residues per protein to yield a brightness of 100 photons/second exceeding the background level. Future work will explore more advanced plasmonic nanostructures to further improve the detected signal per protein.

**Methods**

*Zero-mode waveguide sample fabrication*

A 50 nm-thick layer of aluminum is deposited on cleaned microscope quartz coverslips by electron-beam evaporation (Bühler Syrus Pro 710). In order to ensure the best plasmonic response for the aluminum layer,[28,66] the chamber pressure during the deposition is maintained below $10^{-6}$ mbar and the deposition rate is 10 nm/s. Individual nanoapertures are then milled using gallium-based focused ion beam (FEI dual beam DB235 Strata) with 30 keV energy and 10 pA current.

*Surface passivation*

The ZMW samples are rinsed with ultrapure water and isopropanol and then exposed to oxygen plasma for 5 minutes to remove any remaining organic residues and densify the oxide layer. To protect the aluminum surface and mitigate the corrosion effects,[46,47] the sample is covered by a 5 nm thick polymer layer made of polyvinylphosphonic acid (PVPA, Sigma Aldrich).[49,79] ZMW samples are placed in 2.8 % m/v PVPA solution in water preliminary heated to 90 °C and left for 30 minutes to cover the surface. Then, the samples are rinsed with Milli-Q water and annealed at 80 °C for 10 minutes in a dry atmosphere.

*Protein sample preparation*

β-galactosidase from *Escherichia coli* (PDB 1DP0, UniProtKB P0072, mass 466 kDa) are used as received from Sigma Aldrich without further purification. The stock solution of protein molecules is stored in phosphate buffer solution (PBS, pH 7.4) at -20 °C temperature. Before the measurements, the stock solution is slowly defrosted at 4 °C, then at room temperature (20 °C) and diluted down to 2 µM concentration in a buffer which contained PBS, 1.5 µM pyranose oxidase, 0.83 µM catalase, 10 w/v% D-glucose, 0.5 % Tween20 and 10 mM ascorbic acid at pH 4. D-glucose (≥99.5%), pyranose oxidase (from *Coriolus sp.*, expressed in *E.coli*, M=270 kDa) and catalase (from bovine liver, M=250 kDa) together are abbreviated as PODCAT and play the role of an oxygen scavenger[51] that removes a significant fraction of oxygen from the solution and increases photostability and brightness of β-galactosidase, whereas ascorbic acid is introduced as a protein-friendly antioxidant to reduce pH of the solution for the measurement,[52] and help stabilize the aluminum nanostructures.[49] All the aforementioned chemicals are purchased from Sigma Aldrich. Likewise, p-terphenyl molecules are used as received from Sigma Aldrich, and are diluted in pure HPLC-grade cyclohexane. Emission and absorption fluorescence spectra (Fig. 1d,e) are recorded on an automated cuvette spectrophotometer (Tecan Spark 10M).

*Experimental setup*

The experiments are carried out on a home built confocal microscope with time-resolved fluorescence detection. The Supporting Information Fig. S1 presents a detailed scheme of the setup. The p-terphenyl molecules are excited by a 266 nm picosecond laser (Picoquant LDH-P-FA-266, 70 ps pulse duration, 80 MHz repetition rate). For the β-galactosidase proteins, we use instead a 295 nm picosecond laser (Picoquant VisUV-295-590, 70 ps pulse duration, 80 MHz repetition rate), as the 295 nm excitation yields a slightly better signal to noise ratio than the 266 nm line for this sample. Both laser beams are spatially filtered with 50 µm pinholes, spectrally filtered by a short pass filter (Semrock FF01-311/SP-25) and reflected by a dichroic mirror (Semrock FF310-Di01-25-D) towards the microscope body. For p-terphenyl experiments, the 266 nm laser power is kept at 200 µW except for experiments involving power variation. Similarly, for β-galactosidase experiments, the 295 nm laser power is fixed at 100 µW. The laser powers are measured before the microscope entrance port.

A Zeiss Ultrafluar 40x, 0.6 NA glycerol immersion objective focuses the UV laser beam to a diffraction-limited spot (Supporting Information Fig. S1). A 3-axis piezoelectric stage (Physik Instrumente P-517.3CD) positions an individual ZMW at the laser focus. The fluorescence light is collected through the same objective in an epifluorescence configuration and transmitted through the dichroic mirror (Semrock FF310-Di01-25-D) where it is separated from the laser beam. To further suppress the laser backreflection, a long pass filter (Semrock FF01-300/LP-25) is incorporated in the detection light path. An air-spaced achromatic doublet with 200 mm focal length (Thorlabs ACA254-200-UV) is used as a microscope tube lens to focus the fluorescence light on a 50 µm confocal pinhole. Finally, an emission band pass filter (Semrock FF01-375/110-25) selects the detection spectral range from 310 to 410 nm. A single photon counting photomultiplier tube (Picoquant PMA 175) connected to a photon counting module (Picoquant Picoharp 300) registers the arrival time of each detected photon in a time tagged time resolved mode (TTTR). Fluorescence lifetime measurements feature temporal resolutions of 150 ps at 266nm and 140 ps at 295 nm excitation defined as the full width at half maximum of the instrument response function. All fluorescence traces are analyzed using Symphotime 64 software (Picoquant). The integration time for each trace is 5 minutes.

*FCS analysis*

The FCS correlation data are fitted using a standard three dimensional Brownian diffusion model:[19,53]

$$G(\tau) = \frac{1}{N_{mol}} \left[1 - \frac{B}{F}\right]^2 \left(1 + \frac{\tau}{\tau_d}\right)^{-1} \left(1 + \frac{1}{\kappa^2}\frac{\tau}{\tau_d}\right)^{-0.5} \quad (1)$$

where $N_{mol}$ is the total number of detected molecules, B the background noise intensity, F the total fluorescence intensity, $\tau_d$ the mean diffusion time and $\kappa$ the aspect ratio of the axial to transversal dimensions of the detection volume. Obviously the nanoaperture geometry is more complicated than an open 3D volume, yet this model was found to correctly describe the FCS data inside the nanoapertures,[58,73] when the aspect ratio constant $\kappa$ is set equal to 1. The fluorescence brightness per molecule is then computed as (F-B)/ $N_{mol}$. Figure S3 and S7 shows representative FCS correlation functions and their numerical fits for the different experiments. In all cases, the fit residuals are flat and symmetrically distributed across the zero line, indicating the good quality of the fitting process. Considering the limited signal to noise ratio achieved in the UV, a more elaborate fitting model with a larger number of free components would not bring significant additional knowledge.

*Fluorescence lifetime analysis*

The fluorescence lifetime decay histograms are fitted by means of a Levenberg-Marquard optimization performed on a commercial software (Picoquant SymPhoTime 64). The model performs an iterative reconvolution fit considering the instrument response function (IRF). The region of interest in the temporal decays are set to ensure that more than 96% of all detected photons are considered. The fluorescence decay for p-terphenyl in the confocal reference is fitted with a single exponential function. However, in the case of the nanoapertures, we find that a function with three exponential components provides a better fit to the intensity decay (Fig. 2c). Especially for the smallest 35 nm aperture diameter, the fluorescence decay becomes clearly not single exponential. The reasoning behind the three components is the following: in the case of nanoapertures, we observe that all decays need to take into account a very short lifetime of 10 ps (below the resolution of our system, this lifetime is fixed in the fitting procedure) to interpolate well the initial peak. We relate this contribution to some photoluminescence of the metal or Raman scattering of the sample. We also observe that in order to take well into account the decay tail at long time delays, we have to consider a contribution with fixed lifetime of 0.95 ns corresponding to the lifetime of p-terphenyl in the confocal case. This corresponds likely to a residual fluorescence contribution from molecules lying away from the aperture whose fluorescence emission is not enhanced by the plasmonic nanostructure. Finally, we extract the main contribution from the aperture fluorescence as the lifetime of the last component, which is the only free lifetime in the numerical fit. All the details for the numerical results of the lifetime decay fits are given in the Supporting Information Table S1. We use the same procedure for the fluorescence decay of β-galactosidase in nanoapertures (Supporting Information Fig. S10). Note that due to the large number of tryptophan residues in β-galactosidase, the decay for the confocal reference is also non-single exponential.

*Numerical simulations*

Computations for the electric field intensity distributions shown on Fig. 1c are performed with finite-difference time-domain (FDTD) method using RSoft Fullwave software. The simulated geometry reproduces the experimental ZMW samples characterized by SEM imaging (FEI dual beam DB235 Strata). The complex permittivity for aluminum is taken from the optimized experimental values recorded in ref [66], and the refractive indexes for quartz and water are taken from ref [80]. Each simulation is run with 0.5 nm mesh size and is checked for convergence after several optical periods.

**Supporting Information**

The Supporting Information is available free of charge on the ACS Publications website at DOI:

> Experimental setup, Propagation constant in aluminum nanoapertures, FCS analysis on p-terphenyl molecules, Quadratic dependence of the diffusion time on the decay length, Concentration dependence of p-terphenyl FCS data, Excitation intensity dependence of p-terphenyl fluorescence brightness, FCS analysis on β-galactosidase proteins, FCS analysis on background noise, Comparison of the detection volumes measured with the two molecules, Fluorescence brightness per molecule enhancement of β-galactosidase, Fluorescence lifetime decays of β-galactosidase

**Notes**

The authors declare no competing financial interest.

**Acknowledgments**

The authors thank Antonin Moreau and Julien Lumeau for help with the aluminium deposition. This project has received funding from the European Research Council (ERC) under the European Union's Horizon 2020 research and innovation programme (grant agreement No 723241).

# Supporting Information for

# Deep UV plasmonic enhancement of single protein autofluorescence

# in zero-mode waveguides

Aleksandr Barulin, Jean-Benoît Claude, Satyajit Patra, Nicolas Bonod, Jérôme Wenger*

*Aix Marseille Univ, CNRS, Centrale Marseille, Institut Fresnel, 13013 Marseille, France*

*\* Corresponding author: [jerome.wenger@fresnel.fr](mailto:jerome.wenger@fresnel.fr)*

**Contents:**

## S1. Experimental setup

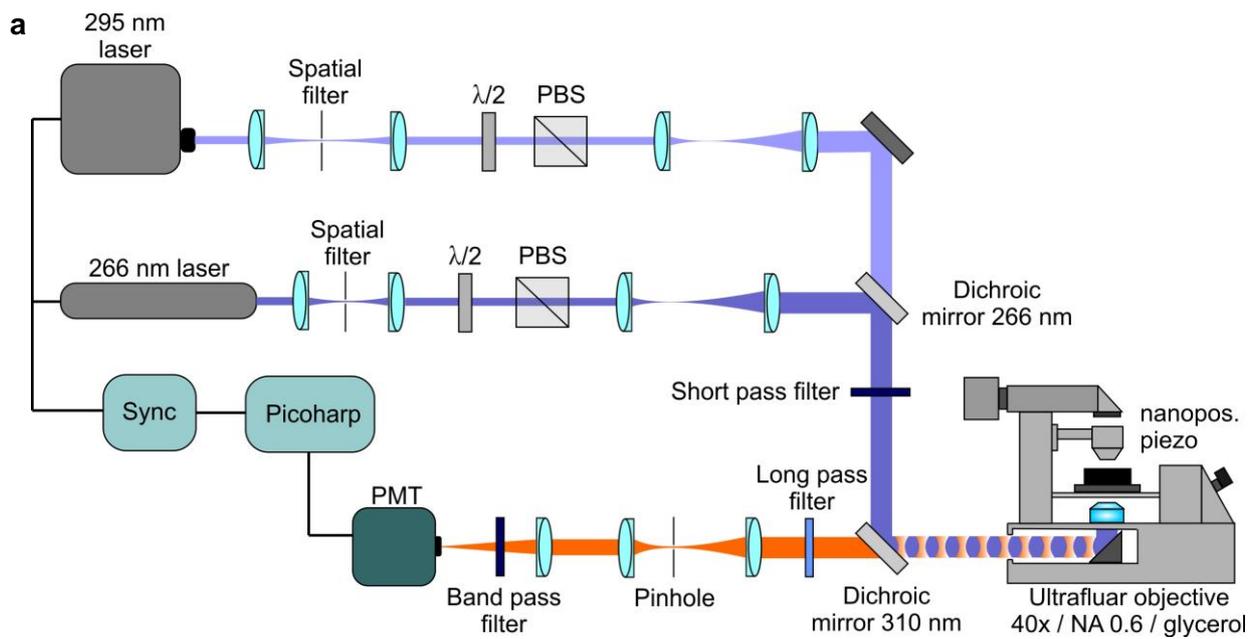

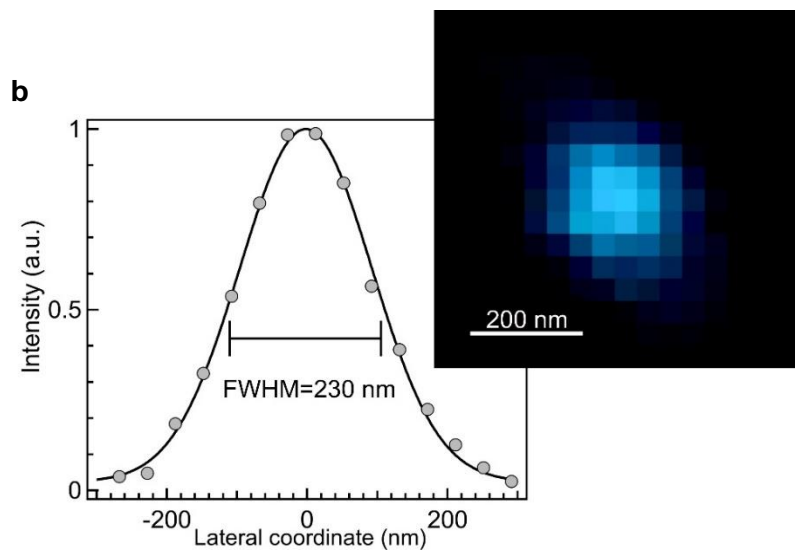

**Figure S1.** (a) Schematic view of the experimental setup. (b) XY scan of a 50 nm ZMW filled with a p-terphenyl solution as a fluorescent emitting source. This image is used to define the microscope point spread function (PSF) whose cross-cut view is shown together with a Gaussian fit.

## S2. Propagation constant in aluminum nanoapertures

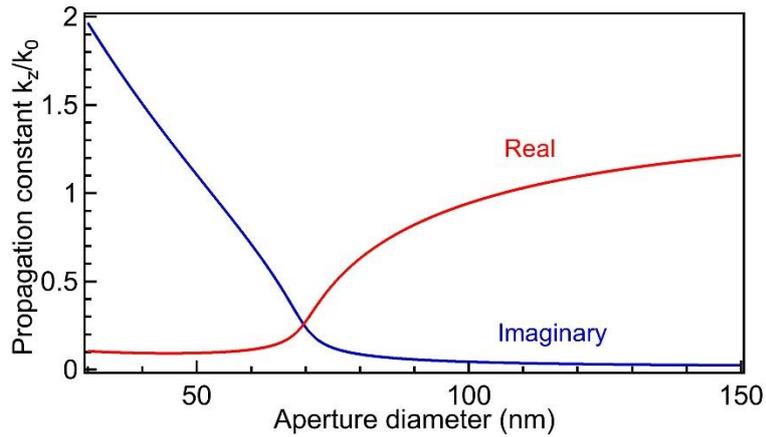

**Figure S2.** Real and imaginary part of the normalized propagation constant $k_z/(2\pi/\lambda)$ along the ZMW main axis as a function of the ZMW diameter. The calculations assume a cylindrical infinite waveguide, and follow a differential method written in a cylindrical geometry.[S1] The aperture is filled with water and the vacuum wavelength is 266 nm. For ZMW diameters below 70 nm, the real part of the propagation constant vanishes and the intensity exponentially decays along the ZMW axis.

## S3. FCS analysis on p-terphenyl molecules

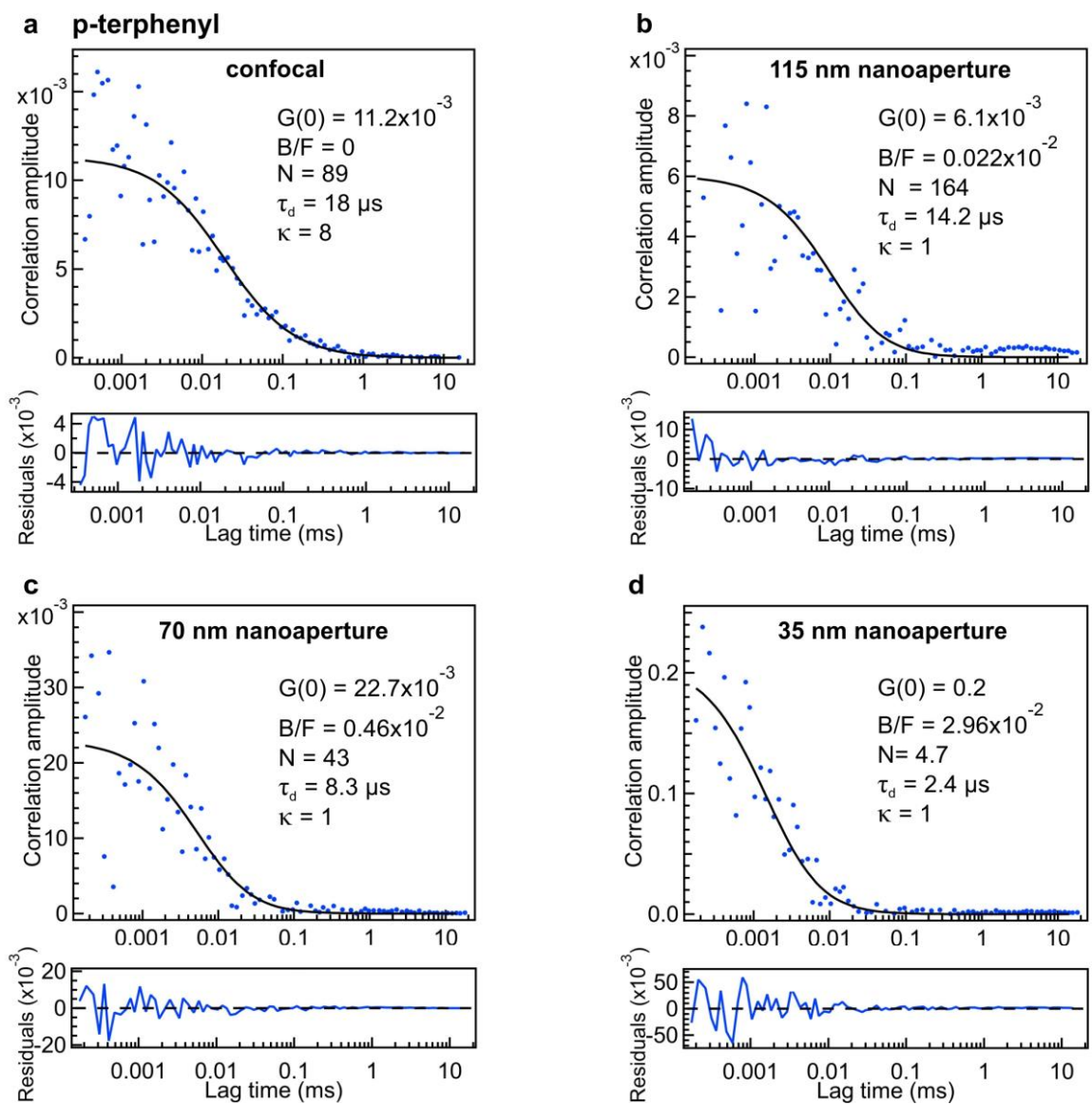

**Figure S3.** FCS correlation functions of p-terphenyl (blue markers) and their numerical fits for the confocal reference (a), as well as in ZMWs of 115 nm (b), 70 nm (c) and 35 nm (d) diameter. The lower traces show the residuals of the fit functions. The fit parameters deduced from the fitting model are shown in each panel. The p-terphenyl concentration used for the ZMWs is 10 µM, and 45 nM for the confocal experiment demonstrated here.

**S4. The FCS diffusion time in the ZMWs has a quadratic dependence with the decay length**

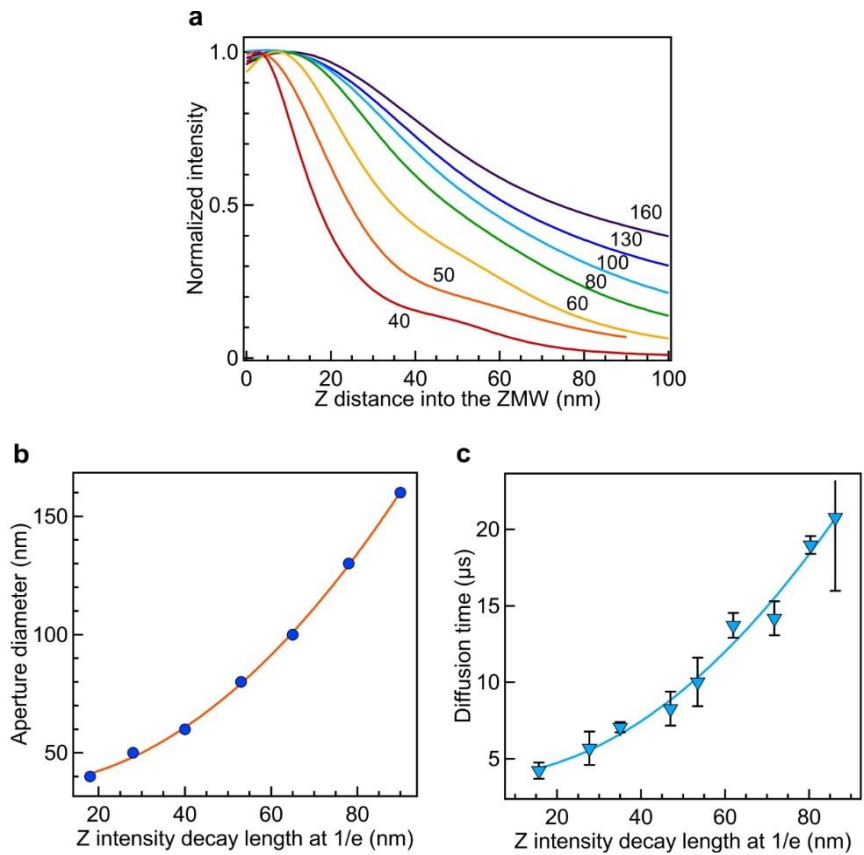

**Figure S4.** (a) Normalized intensity profiles along the center axis of the ZMW computed by 3D FDTD. The numbers on each trace indicate the corresponding ZMW diameter. From the intensity profiles in (a), we determine the characteristic decay length at 1/e. Representing the ZMW diameter (b) or the FCS diffusion time (c) as a function of this characteristic decay length at 1/e, we find a quadratic dependence with the decay length. The lines in (b,c) are numerical fits using a quadratic power law (fixed exponent = 2, the exponent is not a free parameter). The graph in (c) explains the somewhat counterintuitive *linear* dependence of the diffusion time with the ZMW diameter observed in Fig. 2e, and shows that our observations actually follow the *quadratic* dependence expected for Brownian diffusion (which is a supplementary validation). The proper parameter to define correctly the distance dependence in the ZMW diffusion time is the intensity profile decay length, not the ZMW diameter. However, from an end-user point of view, the diameter remains by far the most practical parameter to characterize and compare between the ZMWs, hence our choice for the graphical display in Fig. 2e.

## S5. FCS data on dilutions series of p-terphenyl samples show linear behavior with the concentration

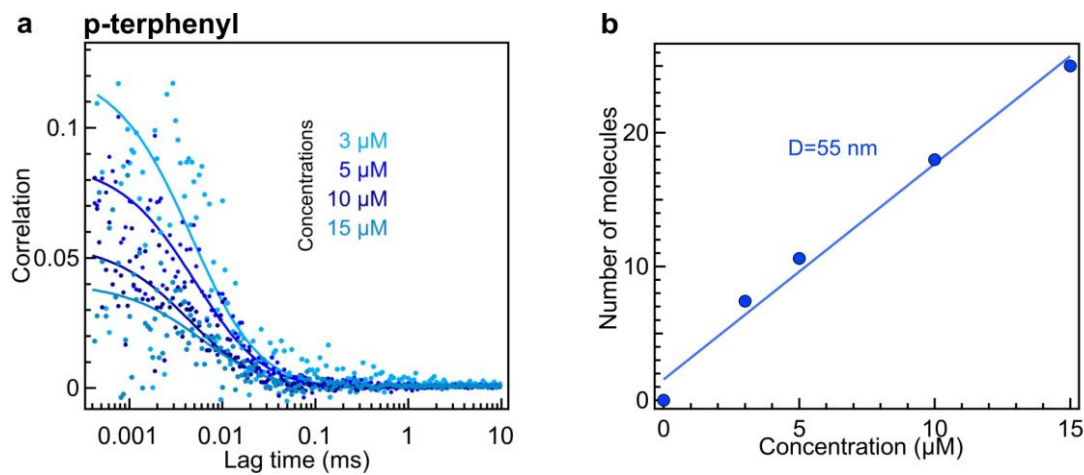

**Figure S5.** (a) FCS correlation functions in a 55 nm ZMW measured at four various concentrations of p-terphenyl. Lower concentrations yield higher correlation amplitude as the FCS amplitude scales with the inverse of the number of molecules. (b) Number of molecules extracted from the FCS correlations versus the sample concentration. The line is a numerical fit. From its slope, we can deduce a volume of 2.8 attoliters (2.8 $10^{-18}$L or 0.0028 µm$^3$) for the 55 nm diameter aluminum ZMW illuminated at 266 nm.

## S6. Excitation intensity dependence of p-terphenyl fluorescence shows linear behavior and no signs for saturation

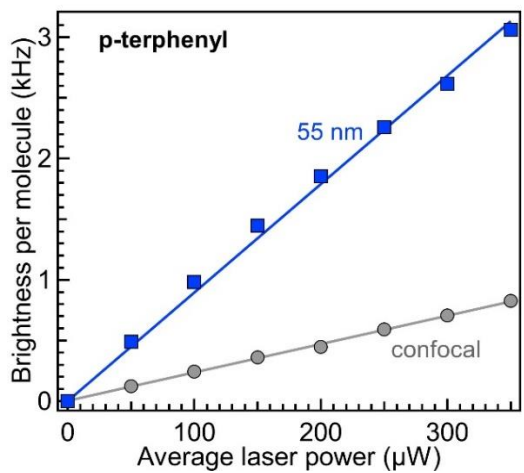

**Figure S6.** Fluorescence brightness per molecule of p-terphenyl in confocal reference and in a ZMW of 55 nm diameter as a function of the 266 nm laser power. The data points follow a linear dependence and do not present any sign of saturation. The brightness per molecule is computed by dividing the average fluorescence intensity by the number of molecules obtained from the FCS fit.

## S7. FCS analysis on β-galactosidase proteins

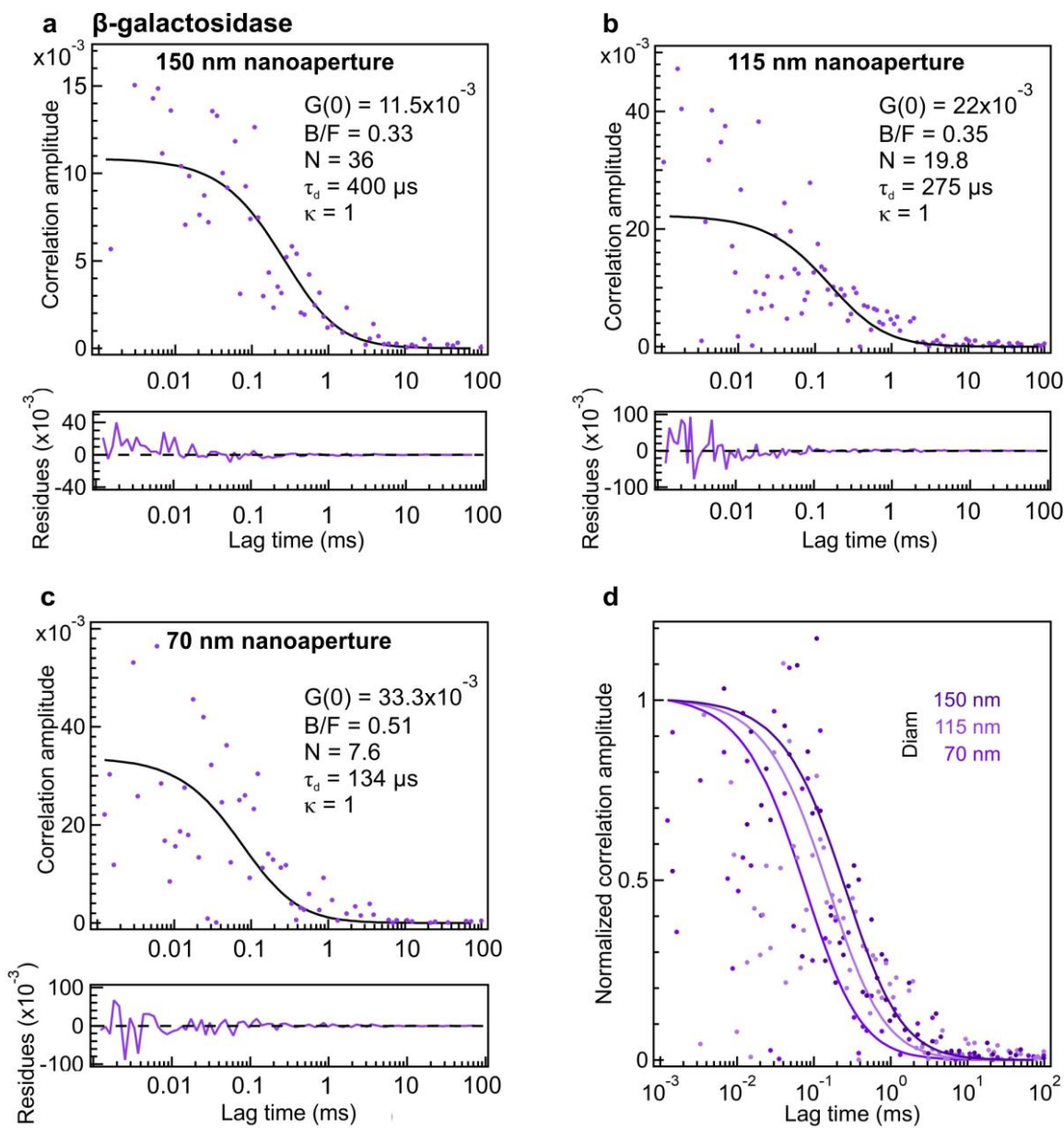

**Figure S7.** FCS correlation functions of β-galactosidase (violet markers) and their numerical fits (black curves) for ZMW diameters of 150 nm (a), 115 nm (b) and 70 nm (c). The lower traces show the residuals from the fit functions. The fit parameters deduced from the fitting model are shown in each panel. The β-galactosidase concentration used for these experiments is 2 µM. (d) Amplitude-normalized FCS correlation functions showing a clear reduction of correlation width as the diameter is reduced.

## S8. FCS analysis on the background noise due to the protein buffer alone does not show any correlation

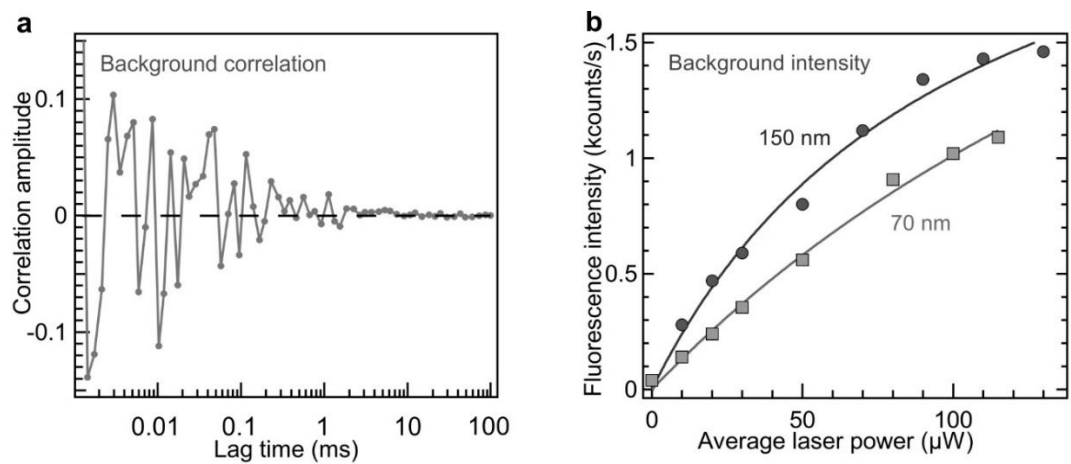

**Figure S8.** (a) Correlation function of a protein buffer in a 70 nm diameter ZMW. The data is symmetric around the zero level indicating no correlation is found. (b) Background fluorescence intensity of protein buffer in the ZMWs versus average laser power. The protein buffer consists of 1.5 µM pyranose oxidase and 830 nM catalase with 10 mM ascorbic acid (pH 4.1).

## S9. The measured ZMW detection volumes show comparable values with p-terphenyl and β-galactosidase

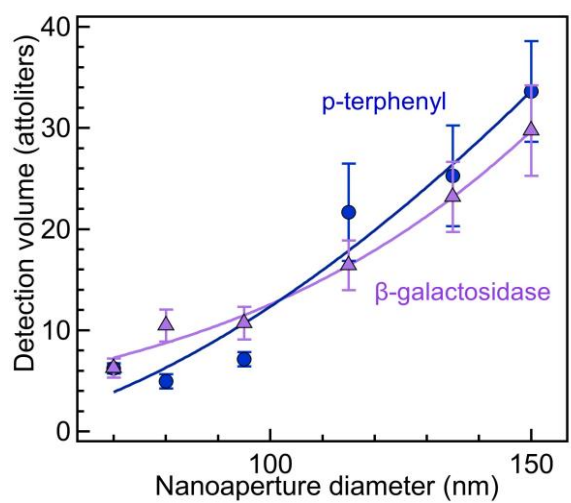

**Figure S9.** ZMW detection volume deduced from the number of molecules and known concentration of the solution as a function of the ZMW diameter. The volumes obtained with β-galactosidase and p-terphenyl coincide, which further validate our results. The lines are guide to the eyes.

## S10. Fluorescence brightness per molecule enhancement of β-galactosidase

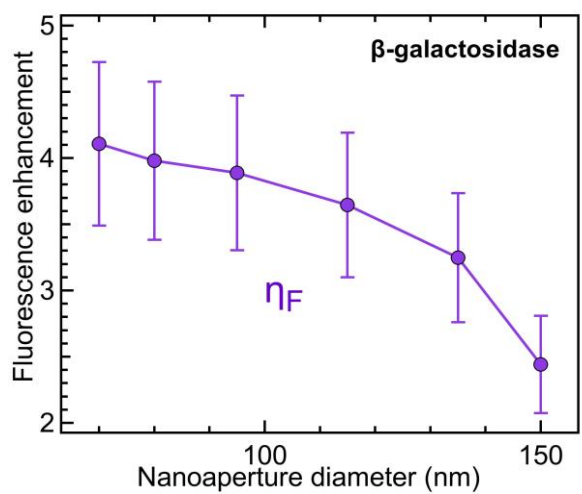

**Figure S10.** Fluorescence brightness per molecule enhancement of β-galactosidase in the presence of ZMWs of increasing diameters measured at 100 µW laser power.

## S11. Fluorescence lifetime decays of β-galactosidase

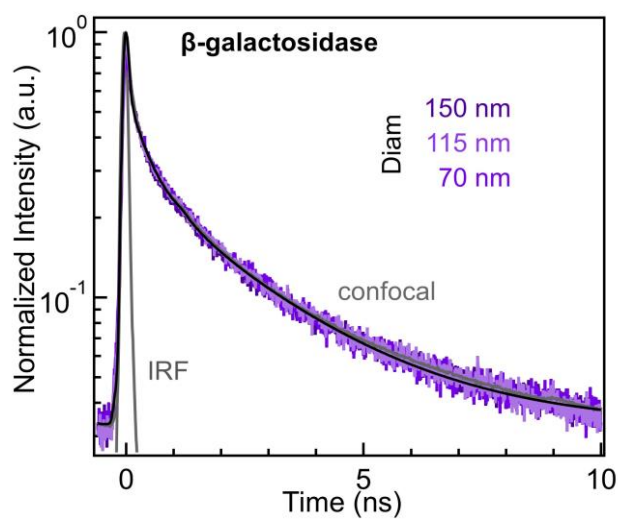

**Figure S11.** Normalized TCSPC lifetime decays of β-galactosidase in free solution and in ZMWs. The normalization was done with fixed intensity at 0.6 ns as a reference point in order to dismiss artefacts of lifetime reduction attributed to Raman scattering or fast lifetime components in the buffer. No visible sign of lifetime reduction is visible here, which indicates that the decay photokinetics of β-galactosidase are largely dominated by its internal non-radiative conversion rate. The fit results are detailed in Table S1.

**Table S1:** Parameters for the triexponential fits of the TCSPC histograms for p-terphenyl in free solution and in ZMWs of different diameters. $\tau_i$ indicate the lifetimes, $I_i$ are the relative intensities of each exponential component and $<\tau_{int}>$ is the intensity-averaged lifetime. 96-99 % of detected photons were included in the region of interest for the fit. For the ZMW fits on p-terphenyl, the first and third lifetime components are fixed at 0.01 and 0.95 ns, respectively. For the ZMW fits on β-galactosidase, only the first component is fixed at 10 ps to account for the residual backscattering of the laser light and the Raman scattered light.

| p-terphenyl | $\tau_1$ (ns) | $\tau_2$ (ns) | $\tau_3$ (ns) | $I_1$ | $I_2$ | $I_3$ | $<\tau_{int}>$ (ns) |
|---|---|---|---|---|---|---|---|
| Confocal ref. | - | - | 0.95 | - | - | 1 | 0.95 |
| 115 nm aperture | 0.01 | 0.62 | 0.95 | 0.04 | 0.38 | 0.58 | 0.78 |
| 70 nm aperture | 0.01 | 0.51 | 0.95 | 0.08 | 0.51 | 0.41 | 0.65 |
| 35 nm aperture | 0.01 | 0.27 | 0.95 | 0.19 | 0.57 | 0.24 | 0.38 |
| **β-galactosidase** | 0.01 | 0.44 | 2.60 | 0.18 | 0.16 | 0.66 | 1.75 |